# Observation of magnetocapacitance in ferromagnetic nanowires

Kulothungasagaran Narayanapillai, Mahdi Jamali, and Hyunsoo Yang[a)]

*Department of Electrical and Computer Engineering, National University of Singapore, 117576, Singapore*

The authors have investigated magnetic domain wall induced capacitance variation as a tool for the detection of magnetic reversal in magnetic nanowires for in-plane (NiFe) and out-of-plane (Co/Pd) magnetization configurations. The switching fields in the capacitance measurements match with that of the magnetoresistance measurements in the opposite sense. The origin of the magnetocapacitance has been attributed to magnetoresistance. This magnetocapacitance detection technique can be useful for magnetic domain wall studies.

[a)] e-mail address: eleyang@nus.edu.sg

Studies of magnetic domain walls (DWs) in magnetic nanowires have attracted attention due to its potential applications in the next generation memories[1] and logic devices[2]. Optical methods such as magneto optical Kerr effect (MOKE)[3] and magnetic soft x-ray microscopy[4] are often used for the study of DW dynamics. Magnetic force microscopy (MFM) is a probe based technique for imaging DW configurations[1]. Scanning electron microscopy with polarization analysis (SEMPA) is an electron based techniques for probing the surface magnetic microstructure of DWs[5]. While each technique has its own advantages, electrical transport measurements are preferred due to their higher temporal and spatial resolutions, leading to direct device applications. Especially, anisotropic magnetoresistance (AMR), giant magnetoresistance (GMR), and tunneling magnetoresistance (TMR) have been extensively utilized to detect magnetic DWs [6,7]. For example, AMR is widely used in Ni enriched alloys for the detection of the magnetic DW dynamical properties including the DW velocity, depinning current, and Walker-breakdown field[7]. Anomalous Hall effect (AHE) is another widely used electrical technique with an out-of-plane magnetic anisotropy, where the DW width is very small[8].

Probe of capacitance provides another method of characterizing the materials and devices. For example, magnetocapacitance (MC) is a phenomenon widely studied in multiferroic materials where the capacitive variation of the multiferroic material in response to external magnetic fields is studied. The magnetic field affects the magnetic ordering and due to the inherent coupling between the ferroelectric and ferromagnetic orders in the multiferroic system, the magnetic signal is reflected in the MC[9]. Spin capacitance occurs due to the accumulation of spin polarized charges at the interface of metals and oxides[10]. In magnetic tunnel junctions (MTJs)[11], the effective interfacial capacitance is due to spin and charge accumulation, and interactions between ions at the interface. The capacitance data also reflect the quality of the interfaces and tunnel barriers in MTJs[12,13]. In this letter, we study MC in ferromagnetic nanowires with in-plane and out-of-plane anisotropy and propose MC as a tool for probing magnetic reversal. Our demonstration of frequency dependent ac-impedance measurements will be useful for the characterization of the magnetic DW devices.

For the in-plane magnetic anisotropy, thin films with the structure of Ta (5 nm)/$Ni_{81}Fe_{19}$ (30 nm)/Ta (5 nm) were deposited on Si/$SiO_2$ substrates by dc-magnetron sputtering with a base pressure of $2\times10^{-9}$ Torr. Semicircular magnetic nanowires with widths of 200 – 900 nm and a diameter of 50 μm were defined by electron beam lithography (EBL) and subsequent argon ion milling. Contact pads were defined by a second EBL step followed by the deposition of Ta (3 nm)/Cu (110 nm) and lift-off process. A top view scanning electron microscopy (SEM) image is shown in Fig. 1(a). In order to nucleate a DW, a bias magnetic field of 750 Oe was initially applied along the *y*–direction and subsequently reduced to zero. Figure 1(b) shows a vortex type DW imaged by MFM (Veeco SPM system) formed at the center of the semi-circular nanowire. A low magnetic moment tip was used for MFM imaging to reduce the interaction between the DW and the tip, and the measurements were performed at a lifted height of 50 nm. Impedance measurements were performed by using an Agilent E4980A precision LCR meter. Standard open and short corrections were performed for calibration to compensate the effect of stray capacitance and inductance from the measurement probes and the coaxial wires. The magnetic field was swept along the *y*–direction during the measurements and ac-impedance measurements were performed for excitation amplitude of 0.125 V and frequency of 500 kHz. The voltage was applied across the $C_1C_2$ ports and measurements were performed across the $P_1P_2$ ports as indicated in Fig. 1(a).

Figure 2(a) shows the resistance versus magnetic field (R-H) measurements for a nanowire with a width of 800 nm. This is a typical AMR behavior[14], which depends on the angle between the local direction of current and the magnetization. When the saturation magnetic field is applied along the *y*–direction, the current is perpendicular to the magnetization. When the external magnetic field decreases from the positive saturation field, the magnetization aligns along the nanowire due to its shape anisotropy leading to the formation of a DW in the center of the nanowire. When the external magnetic field switches to negative polarity and gradually increases to overcome the shape anisotropy, the DW is destroyed, and is seen as the kink points in the curve.

Interestingly, the capacitance versus magnetic field (C-H) measurements in Fig. 2(b) show that the magnetization states clearly affect the capacitive components, as the kink at the point of DW extinction is clearly visible in the capacitance measurements. The capacitance behavior is similar to the inverse of resistance (1/R). We can compare the ratio of MR [=($R_{max}$-$R_{sat}$)/$R_{sat}$], where $R_{max}$ and $R_{sat}$ are the maximum resistance and saturated resistance, respectively, and MC [=($C_{sat}$-$C_{min}$)/$C_{sat}$], where $C_{min}$ and $C_{sat}$ are the minimum capacitance and saturated capacitance, respectively. For example, the MR ratio in Fig. 2(a) is 1.39%, while the MC ratio in Fig. 2(b) is 0.89%. We have also studied the ratio of MR and MC in nanowires with different widths, ranging from 200 to 900 nm, as shown in Fig. 2(c). The MR and MC ratio is relatively constant around 1.45% and 0.9%, respectively, regardless of the width of nanowires. These experiments demonstrate that the capacitive response of the nanowire system can be effectively used for the detection of magnetic reversal and formation of DW.

The vector locus of the resistance ($R$) and reactance ($X$) components from the impedance measurements ($Z = R + jX$) with various frequencies in the range of 50 Hz to 2 MHz show a semi-circle trajectory located in the fourth quadrant for a 600 nm wide nanowire. Absolute value of reactance is plotted in Fig. 2(d) along with the semi-circular fit at zero magnetic field. This confirms that the device behaves like a capacitive system in the measured frequency range. The capacitive behavior of the nanowire system can be understood by the modified Maxwell-Wagner capacitance model[15-17]. This model states that any clustered capacitive system can be modeled by two leaky capacitors in series with one of the leakage components being magnetically tunable. In the present case, the tunable component is the resistive component which arises from AMR.

The equivalent circuit based on the Maxwell-Wagner model is shown in Fig. 3(a), which is divided into parts – the magnetic nanowire and the rest of the measurement path, which includes the line and contacts. The nanowire element can be described by $C_M$, $C_i$, $R_M$, and $R_i$. $C_M$ and $R_M$ are capacitance and resistance, respectively, which depend on magnetic field, while $C_i$ and $R_i$ are field independent capacitance and resistance, respectively. The contact resistance and other parasitic effects can be modeled as shown in Fig. 3(a) into a series resistance ($R_1$) with inductor ($L_1$) which is in series with a parallel capacitor ($C_2$) and resistor ($R_2$), where $R_1$, $L_1$, $C_2$, and $R_2$ are not sensitive to the magnetic field. In order to further quantify the MC, this equivalent circuit has been simplified into two parts; one is depending on the field and the others are not dependent on the magnetic field. The magnetically non-dependent part can be expressed as equivalent impedance $Z_T$ as shown in Fig. 3(b). The magnetically independent part can be removed by deducting the impedances at different magnetic fields mentioned above. Change in the impedance, $\Delta Z$, can be expressed as $\Delta Z = \Delta R + j\Delta X$, where $\Delta R = R_{H=0} - R_{H=500}$, $\Delta X = X_{H=0} - X_{H=500}$.

$$\Delta R = \left( \frac{R_{M0}}{1+4\pi^2 f^2 R_{M0}{}^2 C_{M0}{}^2} - \frac{R_{Mh}}{1+4\pi^2 f^2 R_{Mh}{}^2 C_{Mh}{}^2} \right) \tag{1}$$

$$\Delta X = -2\pi f\left(\frac{C_{M0}R_{M0}^{\ 2}}{1+4\pi^2 f^2 R_{M0}^{\ 2}C_{M0}^{\ 2}} - \frac{C_{Mh}R_{Mh}^{\ 2}}{1+4\pi^2 f^2 R_{Mh}^{\ 2}C_{Mh}^{\ 2}}\right) \quad (2)$$

Here, $R_{M0}$ and $C_{M0}$ are resistance and capacitance at H = 0 respectively, and $R_{Mh}$ and $C_{Mh}$ are resistance and capacitance at 500 Oe. Impedance spectroscopy (IS) was performed from 50 Hz to 2 MHz at two different magnetic fields, 0 and 500 Oe. Figures 3(c) and (d) show $R$ and $X$ components of the IS of a 600 nm wide nanowire at different magnetic fields. The insets show $\Delta R$ and $\Delta X$ with fits. The fitting curves can be derived by Eq. (1) and Eq. (2) as a function of frequency ($f$). Both fittings for $\Delta R$ and $\Delta X$ give comparable fitting values as summarized in Table I.

To further expand the application of this capacitive detection technique for DW studies, we have also tried a system with out-of-plane magnetic anisotropy. Nanowires were patterned with EBL followed by argon ion milling of sputter-deposited thin film having the structure of Ta (4 nm)/Ru (20 nm)/[Pd (0.7 nm)/Co (0.2 nm)]$_{22}$/Ta (4 nm) on a glass substrate. A second photolithography step was used to define Ta (5 nm)/Cu (100 nm) contacts. The width of the nanowire and the Hall bar was defined to be 600 nm and the length of the nanowire is 35 μm as shown in the inset of Fig. 4(b). The vibrating sample magnetometer (VSM) measurements show that the coercivity of the thin film is about 1 kOe as shown in Fig. 4(a). Figure 4(b) shows the anomalous Hall signals across the $C_1C_2$ ports. For the applied field in the $z$-direction, the nanowire shows a square hysteresis with a coercive field of about 2 kOe after patterning.

Figure 4(c) shows the $R$-$H$ measurements across $B_1C_1$ ports with the applied magnetic field in the $z$-direction. The MR reversal process can be understood by electron–magnon scattering processes[18,19]. When the magnetic and anisotropy fields are antiparallel, the anisotropy field tries to maintain the magnetization direction, while the magnetic field destabilizes the magnetization, thus increasing the magnon population. Therefore, in the antiparallel case, the MR linearly increases until the magnetization is switched by the external magnetic field. When the magnetic field and the magnetization are parallel, increasing the magnetic field decreases the magnon population, therefore, the MR linearly decreases with increasing fields as can be observed. In Fig. 4(d) the MC measurements clearly depict the magnetization reversal process and as similar to the NiFe case, the MC measurements show an inverse trend to that of the MR measurements with the same switching fields. It is clear that the MC effect is correlated with the MR effect. The ratio of MR and MC is 0.052% and 0.017%, respectively. Even though the MC effect is small, our observation confirms that MC can be used as a tool to detect magnetization reversal not only in in-plane materials but also in out-of-plane systems.

In summary, we study the magnetocapacitance effect in magnetic nanowires of both in-plane and out-of-plane anisotropy systems. The C-H measurements reveal the same details of the magnetization as that of R-H measurements except that they are in the opposite sense. Based on the Maxwell-Wagner model, we attribute the origin of the MC to the MR effect. These measurements open up the possibility of detecting magnetization reversal and an alternative method to study DW motion.

This work is supported by the Singapore National Research Foundation under CRP Award No. NRF-CRP 4-2008-06.

Figure captions

Fig. 1. (a) SEM micrograph of an 800 nm wide nanowire with electrical leads. (b) MFM image of a vortex domain wall formed at the center of the semi-circular nanowire.

Fig. 2. (a) Resistance of the nanowire under ac-impedance measurements across $P_1P_2$. (b) Capacitance across $P_1P_2$. (c) Magnetoresistance and magnetocapacitance ratio for various widths of nanowires. (d) *R*-*X* plot for the frequency range 50 Hz – 2MHz. The absolute *X* component is plotted with a circular fit.

Fig. 3. (a) Equivalent circuit for the measurement set up with two leaky capacitors representing the nanowire and the rest corresponding to the other effects arising from the coaxial line and contacts. (b) Simplified equivalent circuit with the field dependent components ($C_M$ and $R_M$) and others ($Z_T$). (c) *R* component of impedance spectroscopy (IS) at two different magnetic fields. (d) *X* component of IS. The insets in (c) and (d) show $\Delta R$ and $\Delta X$, respectively, with fits.

Fig. 4. (a) VSM measurements on a Co/Pd multilayer thin film. (b) Anomalous Hall effect measurement. The inset shows a measurement schematic. (c) R-H response of the nanowire across $B_1C_1$. (d) C-H measurements across $B_1C_1$.

TABLE I. Fitting parameters from the $\Delta R$ and $\Delta X$.

| | $\Delta R$ | $\Delta X$ |
|---|---|---|
| $C_{M0}$ (pF) | 558.57 | 562.69 |
| $C_{Mh}$ (pF) | 562.94 | 567.61 |
| $R_{M0}$ (Ω) | 159.56 | 160 |
| $R_{Mh}$ (Ω) | 157.68 | 158.02 |

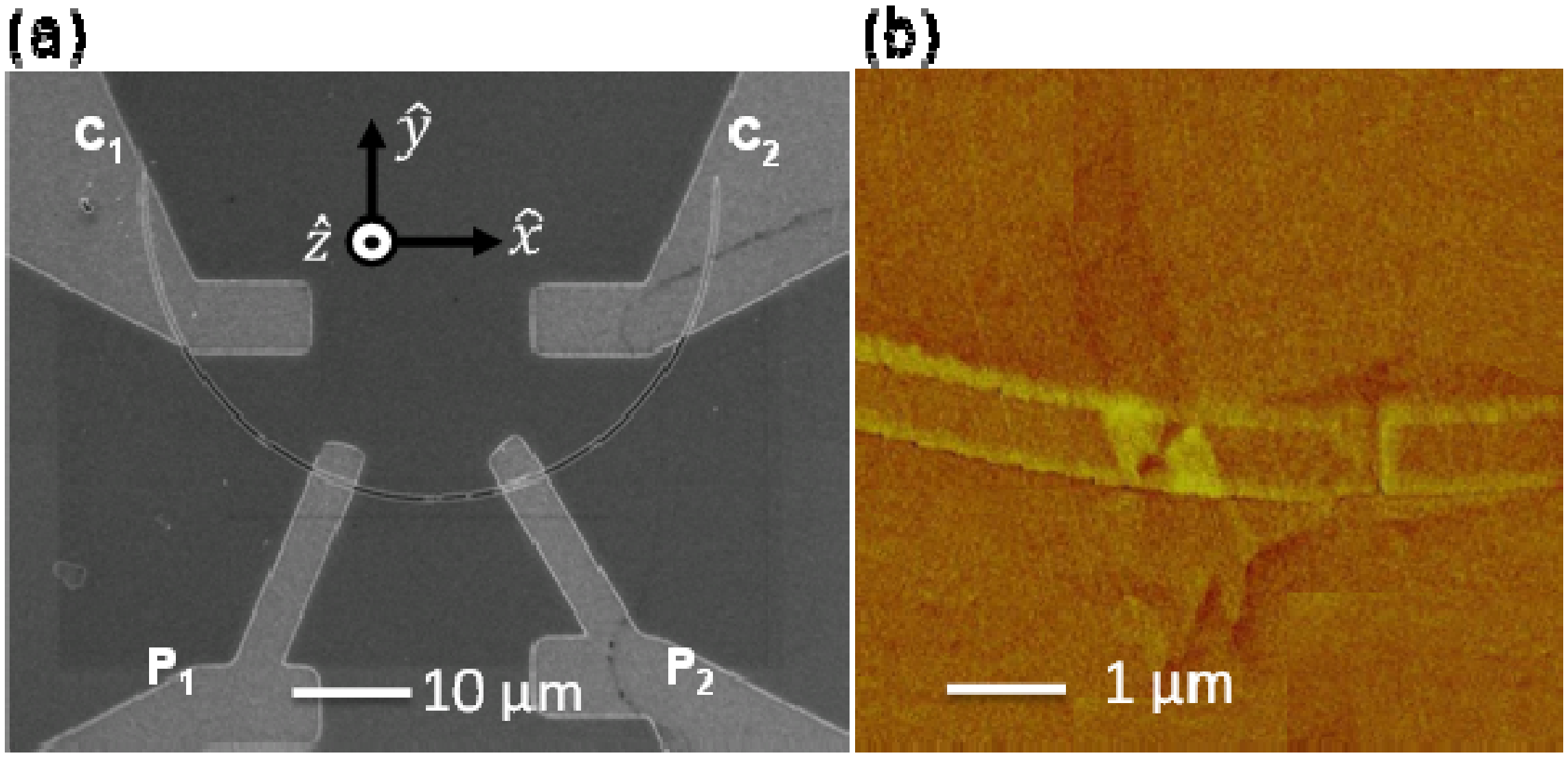

Figure 1

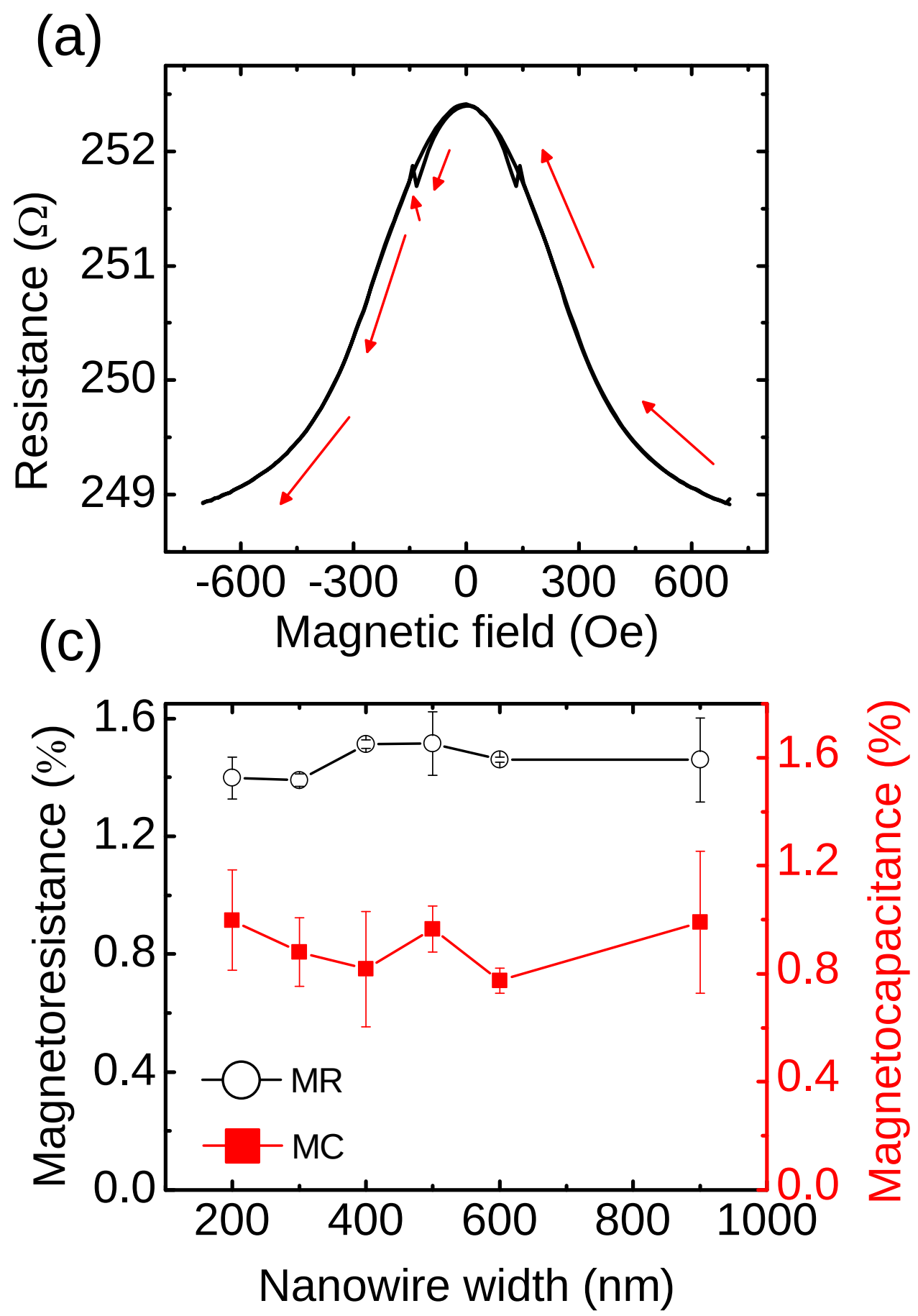

Figure 2

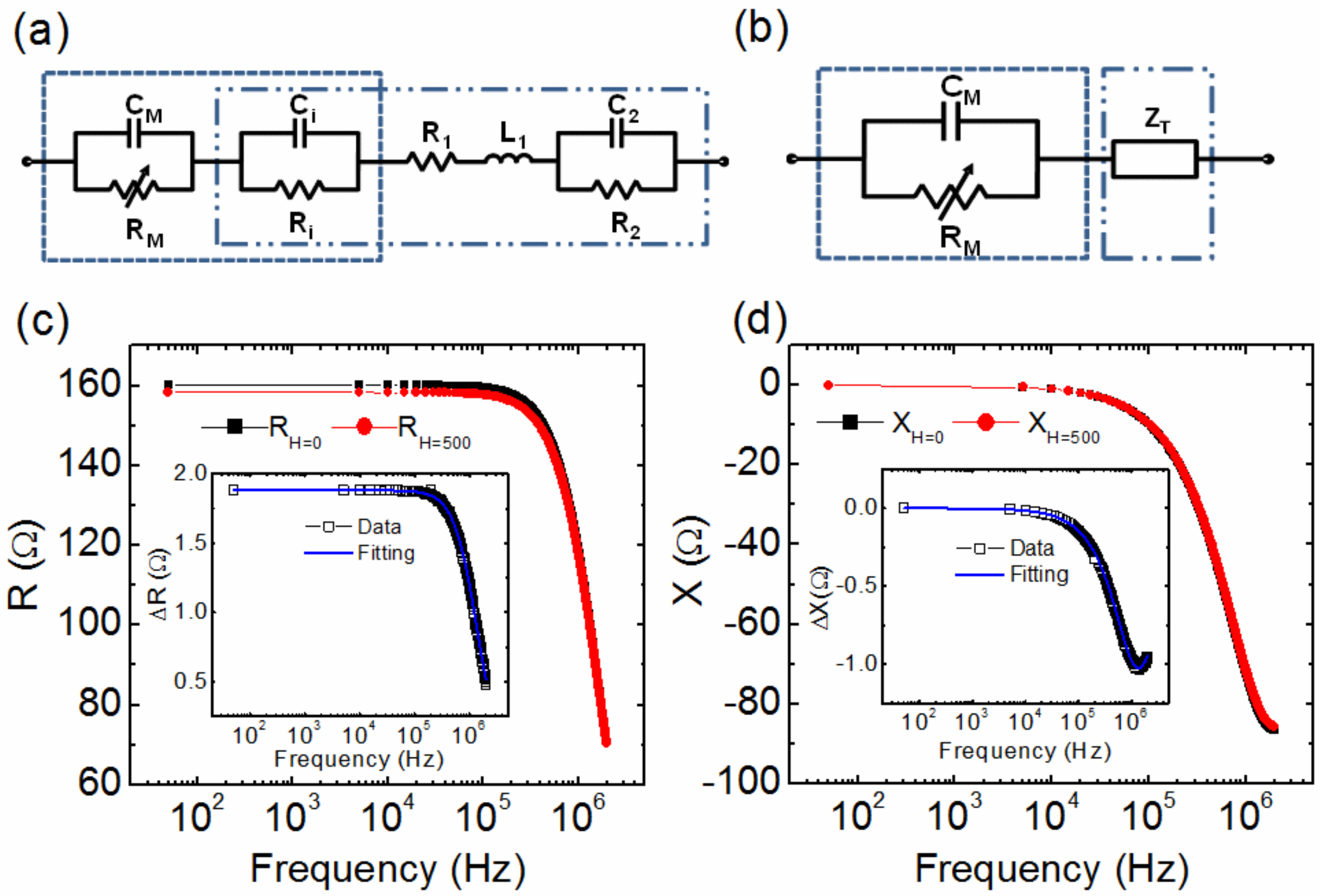

Figure 3

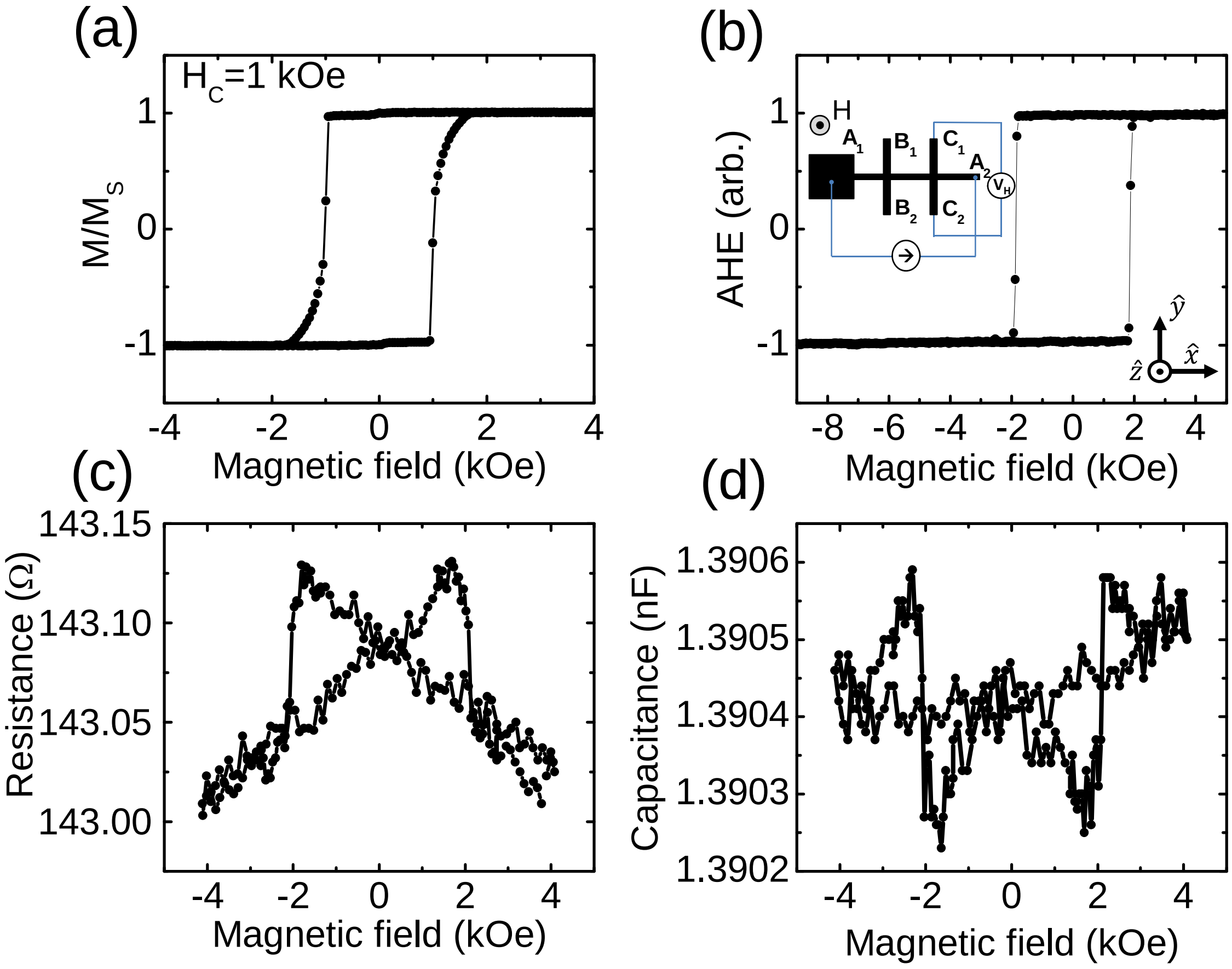

Figure 4